\documentclass[12pt]{iopart}

\usepackage{siunitx}
\usepackage{color}
\usepackage{cite}
\usepackage{amssymb}
\usepackage{graphicx}   
\usepackage{subfigure}
\usepackage{bm}

\usepackage{gensymb}
\usepackage{hyperref}
\hypersetup{
    colorlinks=true,
    citecolor=cyan,
    linkcolor=blue,
    filecolor=magenta,      
    urlcolor=blue,
    }

\begin{document}


\title{Atomistic aspects of load transfer and fracture in CNT-reinforced aluminium}

	\author{Samaneh Nasiri$^1$, Kai Wang$^2$, Mingjun Yang$^2$,\\ Julien Gu{\'e}nol{\'e}$^3$$^,$$^4$, Qianqian Li$^5$, and Michael Zaiser$^1$ }
	
	\address{$^1$WW8-Materials Simulation, Department of Materials Science, Friedrich-Alexander Universit\"at Erlangen-N\"urnberg, 90762 F\"urth, Germany} 
	\address{$^2$School of Materials Science and Engineering, Southwest Petroleum University, Chengdu, Sichuan, PR China}
	\address{$^3$Université de Lorraine, CNRS, Arts et Métiers ParisTech, LEM3, 57070, Metz, France}
	\address{$^4$Labex Damas, Université de Lorraine, 57070, Metz, France}
	\address{$^5$Department of Aeronautics, Imperial College, Prince Consort Road, London SW7 2BZ, UK}
	
	\ead{samaneh.nasiri@fau.de}

\begin{abstract}

This paper describes atomistic simulations of deformation and fracture of Al reinforced with carbon nanotubes (CNTs). We use density functional theory (DFT) to understand the energetics of Al-graphene interfaces and gain reference data for the parameterization of Al-C empirical potentials. We then investigate the load transfer between CNTs and Al and its effect on composite strengthening. To this end, we perform uniaxial tensile simulations of an Al crystal reinforced with CNTs of various volume fractions. We also study the interaction of the embedded CNTs with a crack. We show that the interaction between CNTs and Al is weak such that, under tensile loading, CNTs can easily slide inside the Al matrix and get pulled out from the cracked surface. This effect is almost independent of CNT length and volume fraction. Little load transfer and consequently no crack bridging are observed during the simulation of pristine CNTs threading the crack surfaces. CNTs that are geometrically fixated inside Al, on the other hand, can increase the fracture stress and enhance plastic dissipation in the matrix. CNTs located in front of a growing crack blunt the crack and induce plastic deformation of the Al matrix. Depending on the CNT orientation,  these processes can either increase or decrease the failure stress of the composite. 

\end{abstract}


\newpage
\section{Introduction}
\label{sec:1}
\begin{figure}
	\centering
	\resizebox{0.5\textwidth}{!}{
	\includegraphics{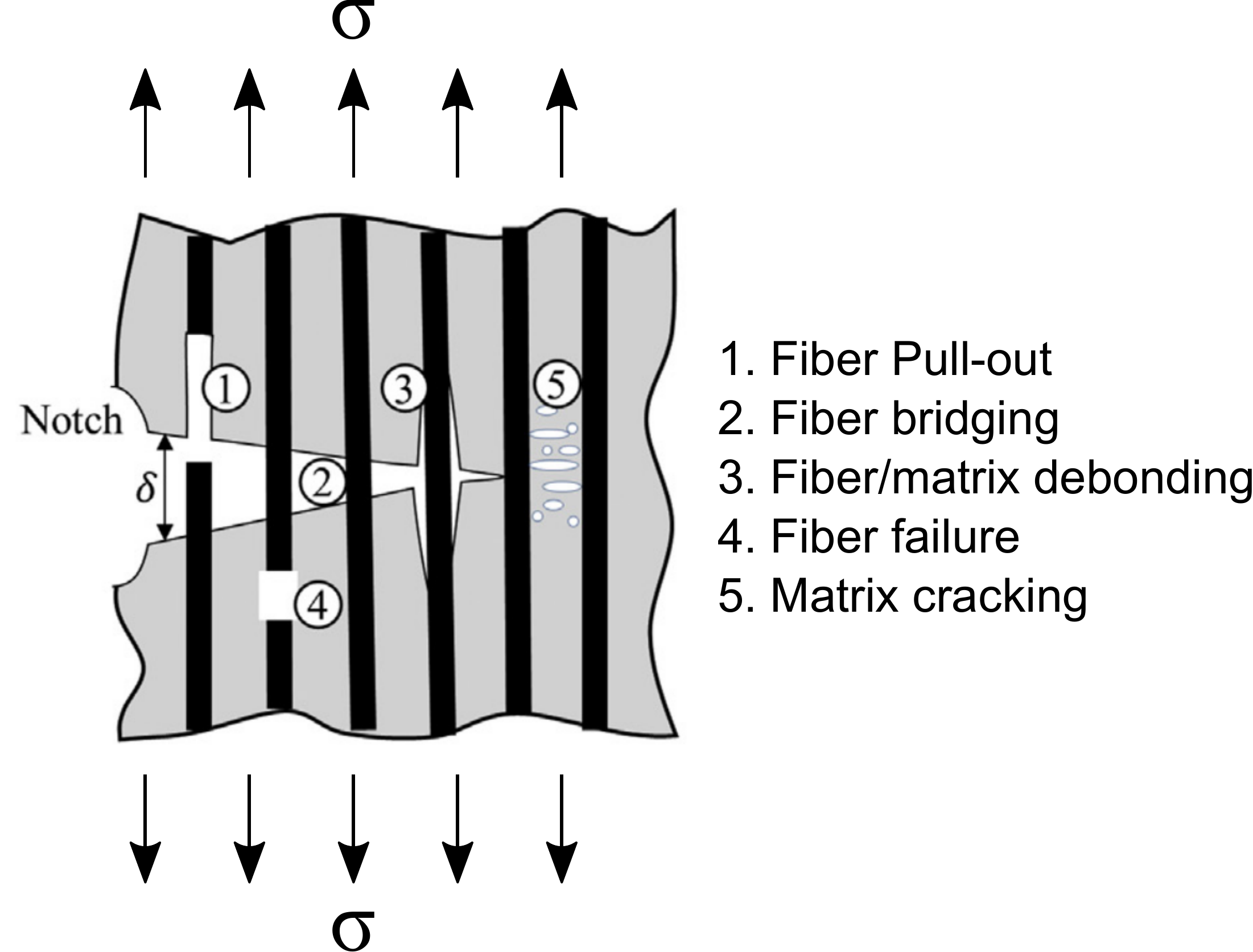}}
	\caption[Fracture modes in fiber reinforced composites]{Suggested fracture modes in fiber/matrix composites \cite{anderson2017-bookfracture}}.
	\label{fig:loadtransfer}       
\end{figure}

Carbon nanoparticles (CNPs), due to their excellent mechanical and structural properties, can significantly improve the mechanical properties of metal or polymer matrices \cite{Zhao2002-PRB, Mielke2004-Chem.Phys.Letter, Sammalkorpi2004-PRB, Khare2007-PRB, Lee2008-science, Yu2000-science, nasiri2016-AIMSmaterialsscience, balandin2011-Nat.Mater, pop2012-MRSbulletin}. Carbon nanotubes (CNTs) exhibit a length-to-diameter ratio (aspect ratio) as high as 1.3e+08 (\cite{zhang2013-ACs, zhang2013-Sci.Rep, stoller2008-Nanoletters}), Young's modulus values of $\sim$ 1 TPa, tensile strength $>$ 100 GPa, and failure strains of 15-30\% depending on chirality for defect-free single-walled CNTs. These properties are incomparable to any other material. Well dispersed CNTs inside metal matrices, therefore, can potentially carry part of the load and increase effective elastic moduli \cite{choi2012-NanoEnergy, Chen2019-Jalcom, Chen2017-Acta, Chen2015-CST, George2005-Scripta, Li2009-PRL, Bakshi2011-carbon}. Experimental studies also showed that embedded CNPs bridge incipient cracks \cite{Chen2015-CST} and impose strong constraints on dislocation motion \cite{kim2013-Nature}, thus potentially increasing yield strength, fracture strength, and possibly also toughness\cite{Li2009-cops.sci.technol, Li2010-compos.sci.technol, Hippmann2013-proc}. Experimental observations indicate that CNTs hinder crack propagation by deflecting the crack from its growth direction. CNTs can also bridge between two crack faces, restrain the crack from opening, and eventually prevent crack growth.  Figure \ref{fig:loadtransfer} depicts the possible mechanisms of crack-CNT interactions. Crack bridging and CNT pull-out phenomena from an Al matrix have been observed experimentally by Chen et al. \cite{Chen2015-CST, Chen2017-Acta} and Park et al. \cite{Park2015-carbon}. Similar observations are documented in several studies on different CNT-reinforced ceramics \cite{Balazsi2006-compB}.

For low-melting lightweight metals such as Al or Mg, the benefit of embedding CNTs is still under discussion since these metals do not wet \(sp^2\) bonded carbon nanoparticles. Experimental and simulation studies reported a small value between 24.0 and 35.0 MPa (\cite{Zhou2016-carbon, Nasiri2019-EPJ}) for the interfacial shear stress (IFSS) between Al matrix and CNTs, indicating a weak interface between Al and pristine CNTs. Therefore, the immersed CNTs into the Al melt tend to bond to themselves rather than Al atoms and agglomerate. At the same time, the interfacial bonding between the nanoparticles and the matrix remains modest after solidification.  In a worst-case scenario, agglomerated nanoparticles with weak bonding into the surrounding metal matrix might act as flaws that deteriorate, rather than improve, the mechanical properties\cite{Li2010-compos.sci.technol}. An extended review of tensile properties of Al reinforced with CNTs can be found in \cite{Jagannatham2020-carbon}.\par

There exist several simulation studies regarding the mechanical behavior of metals with embedded CNTs. Molecular dynamics (MD) studies of CNT-reinforced aluminum composites under uniaxial tensile and compressive loading have been performed by several researchers \cite{Choi2016-JCOMB, Park2018-JMST, Silvestre2014-comscitech, Xiang2017-commatsci,Suk2020-Mater.re.exp}. We find these simulations problematic for several reasons. In some simulations, to model long CNTs, the authors considered CNTs that span the system and cross periodic boundaries in the loading direction. We demonstrate in Section 3.2 that such a set-up is prone to artifacts as it {\em enforces} isostrain loading of matrix and CNT even if no load transfer at the metal/CNT interface is present.  In addition, to define the \(sp^2\) bonded C-C interactions in CNTs, researchers used the Adaptive Intermolecular Reactive Empirical Bond Order (AIREBO) potential \cite{stuart2000-J.Chem.Phys.} which is a reliable potential for simulation of most physical and mechanical properties of hydrocarbons. However, AIRBEO is inadequate to correctly describe C-C bond breaking under tensile loads because the cut-off function of AIRBEO induces an artificial bond stiffening at large bond elongations, which delay bond breakage and induce a ductile failure mode. Several authors have discussed that this is an artifact of the cut-off function \cite{Yang2007-nanotechnology,Pastewka2008-PRB}. In simulations with an improved cut-off function that accounts for changes in the local atomic environment, CNTs found to behave rather brittle at room temperature \cite{Pastewka2008-PRB}. Accordingly, reports of ductile behavior of an Al matrix reinforced by long CNTs may represent artifacts of the potentials and boundary conditions used by the authors. It is also worrying that all the mentioned investigations define the interfacial interactions between the metal and carbon atoms by Lennard-Jones pair potentials which are parameterized by combining Lorentz-Berthelot rules which have been criticized for inaccurate prediction of properties by molecular simulation \cite{Boda2008-MP,delhommele2001-MP}.\par

In the present work, we report the results of a multiscale simulation study of the mechanical properties of Al composites reinforced with CNTs. Firstly, we performed density functional theory (DFT) calculations to establish the energetics of different types of Al-graphene interfaces. The DFT results are then used as reference data for parametrizing the Van der Waals-like potentials for interfacial interactions between Al and \(sp^2\) bonded carbon nanoparticles. We then employed these potentials in molecular dynamics (MD) simulations to establish the strengthening mechanisms in CNT-reinforced aluminum. We investigated the behavior of a periodically continued Al crystal reinforced with single-walled CNTs(5,5) under uniaxial tensile load, aiming to understand the effect of CNT length and volume fraction on tensile strength of the Al-CNT composite. To study the interaction of cracks with CNTs under Mode I loading, we considered an Al crystal containing a CNT and an edge crack. We investigated crack deflection and crack bridging mechanisms by CNTs in Al by changing the CNT length, position, entanglements, and angle of the CNT with respect to the crack front.

\section{Density-functional calculations: Methods and results}
\label{sec:2}
\begin{figure*}
	\centering
	\resizebox{0.8\textwidth}{!}{
	\includegraphics{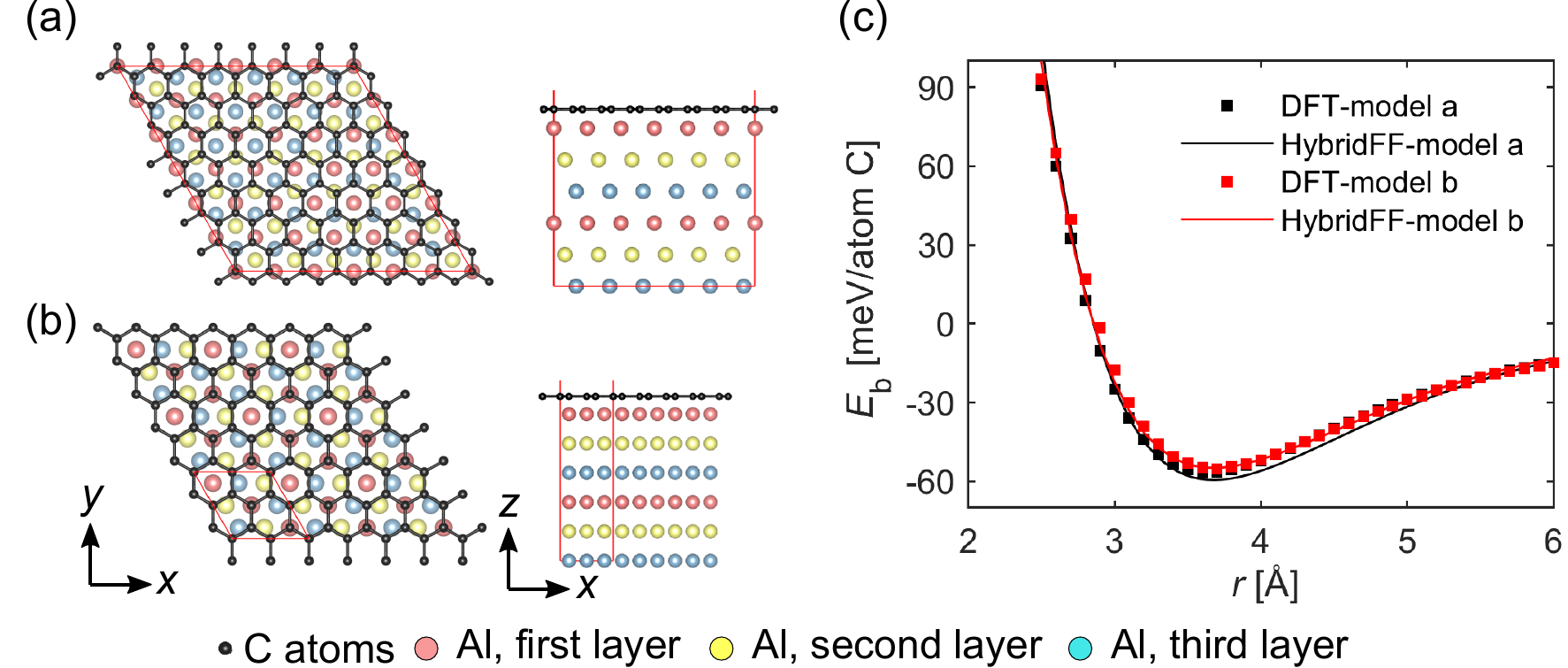}}
	\caption{(a) and (b) top views (left) and side views (right) of the geometry for model a and model b respectively. (c) Al-graphene binding energy for both models calculated with DFT (dotted data) and empirical potentials (Morse potential for Al-C interactions)(lines).}
	\label{fig:1}       
\end{figure*}

We performed DFT calculations to parametrize the interatomic potentials between Al and $sp^2$ bonded carbon. To this end, we considered a planar geometry amenable to periodic continuation containing six-layer Al(111) slabs adhering to monolayer graphene (single-walled CNT with an infinite diameter). Both Al slabs and graphene are oriented parallel to the $xy$ plane of the Cartesian coordinate system. Periodic boundary conditions are applied in all three spatial directions with a 20 \AA{} vacuum gap between the periodic replicas in the $z$ direction. Two variants of this structure are constructed as shown in Figure \ref{fig:1}. For model (a), p \((7 \times 7)\) graphene and p \((6 \times 6)\) Al slabs were used in calculations. For model (b), we used p \((2 \times 2)\) graphene with three atoms per Al layer as in the work of Gong et al. \cite{gong2010-J.Appl.Phys.} and Khomyakov et al. \cite{khomyakov2009-PRB}. Model (a) approximates an incoherent interface, while model (b) approximates a coherent interface. In both models, the lattice parameters were set to their experimental values, and the metal surface was strained to fit the graphene supercell.

The DFT calculations were carried out by the Quantum Espresso package version 6.1 \cite{giannozzi2009-J.Phys.Condens.Matter}, employing PBE-based projected augmented wave(PAW) potentials \cite{blochl1994-PRB}. Previous studies have shown that the optB88-vdW \cite{becke1988-PRA} exchange-correlation functional gives a reasonably accurate prediction of both interlayer distance and binding energy for graphite \cite{graziano2012-J.Phys.Condens.Matter}, so here we choose the optB88-vdW functional in all our DFT calculations. In our calculations, we used the kinetic energy cutoffs of 40 Ry and 450 Ry for the wave function and charge density calculations, respectively. A Methfessel-Paxton smearing of 0.01 Ry was used for the electronic convergence, and the convergence for self-consistency calculations was less than 0.0001 Ry. In the calculations of model (a), only the gamma k-point was used due to the large supercell, while an \(8\times8\times1 \) k-points grid was used in the calculations of model (b). For additional verification of the method, we simulated the same configuration (hpc-fcc) as in the DFT calculations by Christian et al. \cite{Christian2017-carbon} of graphene adhering to an Al(111) slab and found good agreement with their work.\par

Since our aim is to obtain data for parameterizing an interaction potential, rather than performing an accurate analysis of the interface structure and energy using DFT, we do not carry out any structural relaxation but simply separate the graphene sheet rigidly from the Al(111) slab without geometry relaxation of either the Al slab or the graphene. We evaluate the binding energy per carbon atom $E_\mathrm{b}$ as
\begin{eqnarray}
E_\mathrm{b} = [E_\mathrm{tot} - (E_\mathrm{Al} +E_\mathrm{gr})]/n \label{eq1} 
\end{eqnarray}

$E_\mathrm{tot}$ is the total energy of the graphene-Al(111) system, $E_\mathrm{Al}$ represents the energy of the six-layer Al(111) slab, and $E_\mathrm{gr}$ and $n$ represent the energy of the free-standing graphene sheet and the number of carbon atoms in the graphene sheet, respectively. As depicted in Figure \ref{fig:1}, Al manifest weak interaction (physisorption) interfaces with graphene sheets with adsorption energy of around 0.03--0.05 eV/carbon atom, and equilibrium interfacial distances larger than 3.0 $\si{\angstrom}$ \cite{Vanin2010-PRB, gong2010-J.Appl.Phys.}. In these interfaces, the energy differences between different conformations of the graphene on top of the Al slab are minor and hence the energy 'landscape' is rather flat.

\section{Molecular simulations}
\label{sec:3}
The atomistic simulations presented in this work were performed with the molecular dynamics code LAMMPS \cite{LAMMPS} (version 16 Mar 2018). Atomsk package \cite{Atomsk}, and VNL builder in QuantumATK \cite{schneider2017-iop,Stradi2017-iop} were used to create the initial atomic configurations. Data visualization and post-processing were obtained with the help of the Open Visualization Tool (OVITO) \cite{ovito}. Common neighbour analysis (CNA) \cite{Faken1994-comput.mat.sci, Honeycutt1987-J.Phys.Chem} was used to analyze the crystallographic structure around atoms and to identify defects and dislocations in the Al matrix.

\subsection{Interaction potentials}

The chosen interatomic potential is based upon combining potentials of different types that are separately optimized to describe the energetics and mechanical properties of covalently bonded carbon, and of aluminum. We usend Embedded Atom model (EAM) potential of Mishin et al. \cite{mishin2004-acta} for describing Al-Al interactions and the modified screened reactive empirical bond-order (scr-REBO) potential by Pastweka et al. \cite{Pastewka2008-PRB} for C-C interactions. The scr-REBO potential uses a local environment-dependent cut-off function that leads to smooth changes in force and a much better description of bond-breaking and re-forming comparing to conventional truncated REBO potentials. The weak interactions in a physisorbed interface between Al and $sp^2$ bonded carbon can be described by a Morse potential as in the approach used by \cite{Moseler2010-ACS} for simulating Ni-graphene interactions. We used a Morse potential of the form $E_{\mathrm{M}} = D_{\mathrm{M}} (\exp[-2\alpha (r - r_0)] - 2 \exp[-\alpha (r - r_0)])$  for Al-C which has been parametrized using our DFT calculations and fitted to represent the physisorbed interface state between graphene and Al, where $r$ is the bond distance, $r_0$ the equilibrium bond distance, $D_\mathrm{M}$ is the well depth, and $\alpha$ controls the stiffness of the potential. To parametrize the Morse potential, we used the same configuration as in the DFT simulations, i.e., we separate the graphene sheet rigidly from the Al(111) surface, which allows us to directly compare the MD and DFT energies (Figure \ref{fig:1}). The resulting parameters for the Al-C Morse potential are $D_\mathrm{M}$ = 0.0048 eV,  $r_{0}$ = 4.52 $\mathrm{\AA}$ and $\alpha$ = 1.0 $\mathrm{\AA}^{-1}$.  We emphasize that a combination of structurally dissimilar potentials --- EAM, scr-REBO, and Morse --- as used here can produce adequate results only as long as the interactions described by the respective terms can be considered as additive. That is approximately true for physisorption governed by van der Waals forces, which is not expected to significantly alter the bonding characteristics of either metals or covalently bonded carbon. On the other hand, situations where chemical bonds are established at the metal-carbon interface or where reactive processes or mechanical mixing occur, cannot be adequately described in this manner. In particular, we note that our approach {\em a priori} disallows carbide formation.\par

\subsection{Uniaxial tensile deformation of CNT-reinforced Al composites}
\label{subsec:3.1}

This section studies the behavior of a periodically continued Al crystal reinforced by a single-walled CNT(5,5) under uniaxial tensile loading, aiming to understand the effect of length and volume fraction of CNTs on the tensile strength of Al-CNT composite.

\subsubsection{Sample preparation}
\label{subsec:3.1.1}

We created a monocrystalline Al block with faces aligned with the [100], [010], and [001] directions of an fcc crystal lattice. Al atoms are initially located on the sites of a perfect fcc lattice. The simulation box is periodic in all spatial directions. The energy of the system is minimized through structural relaxation using a CG algorithm. The system is accepted as relaxed when both the change of energy between successive iterations divided by the energy magnitude is less than or equal to the tolerance ($10^{-8}$) or the 2-norm (length) of the global force vector is less than the threshold value of $10^{-8}$ eV/$\mathrm{\AA}$ as in \cite{vaid2019-Materialia}. A single-walled CNT(5,5) is then embedded into the center of the Al block, such that its longitudinal axis is parallel to the [010] lattice direction, which aligns with the $y$ direction of the Cartesian coordinate system. The embedded CNT has either equal length as the periodic cell (i.e., it extends across the periodic boundary) denoted as "long CNT"  or it has a 2.46 $\si{\angstrom}$ shorter length than the Al matrix such that it terminates within the periodic cell denoted as "short CNT" in Figure \ref{fig:alc1}. To create the requisite space between Al and C atoms, we removed the Al atoms that were located inside the CNT or within a distance of less than 3.0 $\si{\angstrom}$ from C atoms. To equilibrate the system, Al atoms around the CNT were annealed at 1500K (i.e., above the melting temperature of Al) for 20 ps in the $NpT$ ensemble at zero pressure by using the Nos\'e-Hoover thermostat and barostat, then quenched to 0.1K at a rate of $10^{-12} K.ps^{-1}$. During the melt-quench cycle, the boundary atoms were fixed on the sites of the original crystal lattice. In this case, crystallization starts from the boundary and leads to a monocrystalline Al block surrounding the CNT, which remains aligned with a [110] lattice direction. Carbon atoms are fixed during this process. A final structural minimization is then performed to remove any excess energy in the system. We then calculated the Al/CNT interface energy (\(E_{\mathrm{int}}\)) as $E_{\mathrm {int}} = [E_{\mathrm{Al-CNT}} - (E_{\mathrm{Al-void}} +E_{\mathrm{CNT}})]/n$, where $E_{\mathrm{Al-CNT}}$ is the total energy of the Al-CNT system, $E_{\mathrm{Al-void}}$ represents the energy of the Al block with a cylindrical void in the middle induced by the CNT, and $E_{\mathrm {CNT}}$ and $n$ represent the energy of the free-standing CNT and the number of carbon atoms in the CNT, respectively. The calculated interface energy for the minimized structure is equal to $E_{\mathrm{int}}$ = 0.053 eV/carbon atom that matches well to the equilibrium binding energy from our DFT calculations (Figure .\ref{fig:1}). \par

\begin{table*}
	\centering
	\caption[Characteristic of the simulation for length effect calculations]{Geometrical parameters of the simulated samples for investigation of CNT length effect on composite strength. The unit of the length and diameter is $\si{\angstrom}$.}
	\begin{tabular}{llllllll}
		\hline\noalign{\smallskip}
		Samples&$L_x$ &$L_z$& $L_y$&$D$& $f_\mathrm{CNT}$[\%] \\
		\noalign{\smallskip}\hline\noalign{\smallskip}
		1&113.40&113.40&101.25&6.78&0.27\\
		2&113.40&113.40&202.50&6.78&0.27\\
		3&113.40&113.40&303.75&6.78&0.27\\
		4&113.40&113.40&405.00&6.78&0.27\\
	\end{tabular}	
	\label{tabl:LE}
\end{table*} 

\begin{table*}
	\centering
	\caption[Geometrical parameters of the simulated samples for volume fraction effect calculations]{Geometrical parameters of the simulated samples to study the effect of CNT volume fraction on composite strength. The unit of the length and diameter is $\si{\angstrom}$.}
	\begin{tabular}{llllllll}
		\hline\noalign{\smallskip}
		Samples&$L_x$ &$L_z$& $L_y$& $D$ &$f_\mathrm{CNT}$[\%] \\
		\noalign{\smallskip}\hline\noalign{\smallskip}
		1&113.40&113.40&101.25&6.78&0.27\\
		2&68.85&68.85&101.25&6.78&0.76\\
		3&56.70&56.70&101.25&6.78&1.10\\
		4&44.55&44.55&101.25&6.78&1.72\\
		5&28.35&28.35&101.25&6.78&4.40\\
	\end{tabular}	
	\label{tabl:VE}
\end{table*}

\subsubsection{Simulation set-ups and results}
\label{subsec:3.1.2}

Uniaxial tensile tests are performed by stretching the simulation box with a constant strain rate of 0.001 $\mathrm{ps^{-1}}$ parallel to the $y$ direction. Temperature is kept at 0.1 K, and cross-contraction of the simulation box is allowed in both lateral dimensions by enforcing zero pressure using a Nos\'e-Hoover thermostat and barostat in $NpT$ ensemble. We studied the effect of CNT length by systematically varying the size of the simulation box in the loading direction ($L_y$) for both short and long CNT configurations (Table \ref{tabl:LE}). In this set of simulations, the CNT volume fraction $f_\mathrm{CNT}$ is kept at a constant value of 0.27\%. The effect of CNT volume fraction on composite strength is then studied by varying the lateral sizes ($L_x$, and $L_z$) of the system according to Table  \ref{tabl:VE} while keeping the $L_y$ fixed.  

\begin{figure*}
	\centering
	\resizebox{0.95\textwidth}{!}{
	\includegraphics{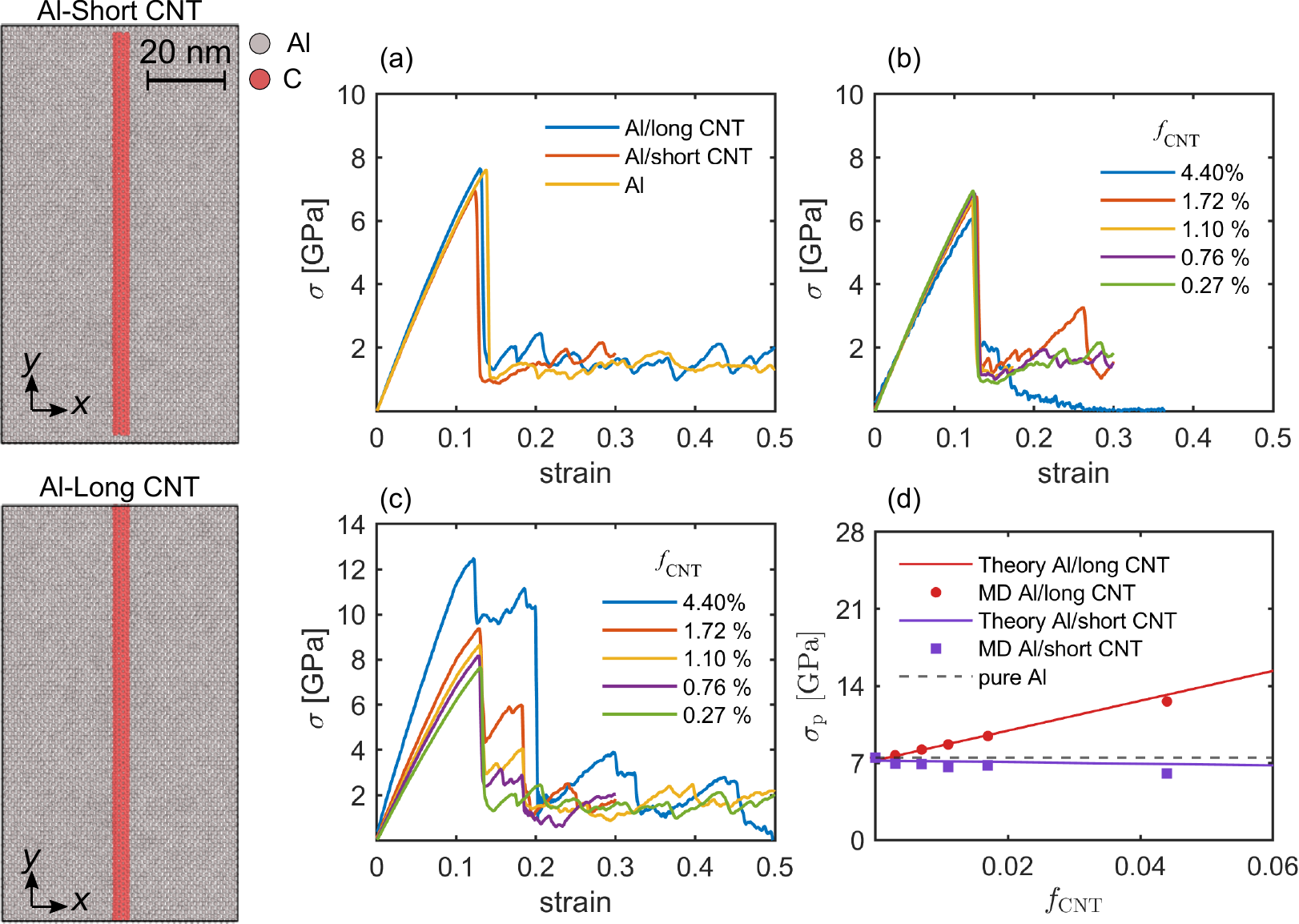}}
	\caption[Atomic configurations of "short" and "long" CNTs embedded inside a single crystal Al]{Left: atomic configurations of "short" and "long" CNTs embedded inside an Al single crystal. Right: (a) Comparison between the stress-strain curves of an Al single crystal and Al-CNT composites  with CNT volume fraction $f_\mathrm{CNT}$ = 0.27 \%.  (b) Influence of the CNT volume fraction $f_\mathrm{CNT}$ on composite strength for "short" CNTs and (c) for "long" CNTs.  (d) Dependency of peak stress on CNT volume fraction for both long and short CNTs.}
	\label{fig:alc1}       
\end{figure*}

Figure \ref{fig:alc1} (left) shows the initial configurations of both short and long CNTs in the Al matrix, whereas  Figure \ref{fig:alc1} (a-c) shows stress-strain curves compared with that of an Al single crystal. In the case of Al reinforced with short CNTs, the composite strength $\sigma_\mathrm{p}$, defined by the highest value of stress reached on the stress-strain curve, is slightly lower than the Al single crystal strength  $\sigma_\mathrm{Al}$ as shown in Figure \ref{fig:alc1} (a). According to Figure \ref{fig:alc1} (b) and (d), increasing the volume fraction $f_\mathrm{CNT}$ of "short" CNTs leads to a further decrease in strength. As we observed from our DFT calculations, the interface between Al and CNT is weak - the interface energy is low, and the interface shear strength (IFSS) is of the order of 30 MPa or less \cite{Nasiri2019-EPJ, Zhou2016-carbon}. Therefore, when the Al-CNT system is loaded in tension, the pristine short CNT aligned parallel to the loading direction can easily slide inside the matrix without carrying a significant part of the applied load. Therefore, under the axial load, the Al atoms at the Al/CNT interface are stretched, while the short CNT is sliding inside the Al and are not stretched. Consequently, the short CNT acts like a void that is elongated along the loading direction, leading to an approximate decrease in composite strength $\sigma_\mathrm{p}$, the peak stress in the stress-strain curve, like

\begin{equation}
\sigma_\mathrm{p} = (1-f_\mathrm {CNT}) \sigma_\mathrm{Al}
\end{equation}

as shown in Figure \ref{fig:alc1} (d), where  $\sigma_\mathrm{Al}$ is Al failure strength. \par

In the case of Al reinforced with long CNTs, we observed an increase in the composite strength as we increase the CNT volume fraction as shown in Figure \ref{fig:alc1} (c-d). Upon loading, one observes two load drops in Figure \ref{fig:alc1}(c), first at a strain $\approx 0.13$ when the Al matrix undergoes profuse dislocation nucleation ($\epsilon_\mathrm{Al}^\mathrm{f}
$), and then again at a strain $\approx 0.2$ when the CNT ruptures ($\epsilon_\mathrm{CNT}^\mathrm{f}$). We should emphasize that the increase in the composite strength is not a consequence of any load transfer at the Al/long CNT interface but a direct result of the applied boundary conditions enforcing isostrain deformation of the Al block and the embedded CNT. The isostrain deformation, before the first load drop, forces both Al and CNT to deform in parallel. Thus the peak stress $\sigma_\mathrm{p}$ fulfills the simple relationship 

\begin{equation}
\sigma_\mathrm{p} = \epsilon_\mathrm{Al}^\mathrm{f} [(1-f_\mathrm{CNT}) Y_\mathrm{Al} + f_\mathrm{CNT}Y_\mathrm{CNT}]
\end{equation}

as demonstrated in Figure  \ref{fig:alc1}(d) where $Y_\mathrm{Al}$ and $Y_\mathrm{CNT}$ represent Young's modulus of Al and CNT, respectively. However, this finding is physically doubtful since, as demonstrated by our simulation of "short" CNTs, even a very small change of the CNT configuration at the boundary of the simulation supercell suffices to destroy the enforced isostrain deformation and the strengthening effect which, in consequence, must be deemed a simulation artifact caused by the improper choice of boundary conditions. 

\subsection{Crack propagation in CNT-reinforced Al crystals under tensile loading}
\label{subsec:3.2}

\begin{figure*}
	\centering
	\resizebox{0.8\textwidth}{!}{
	\includegraphics{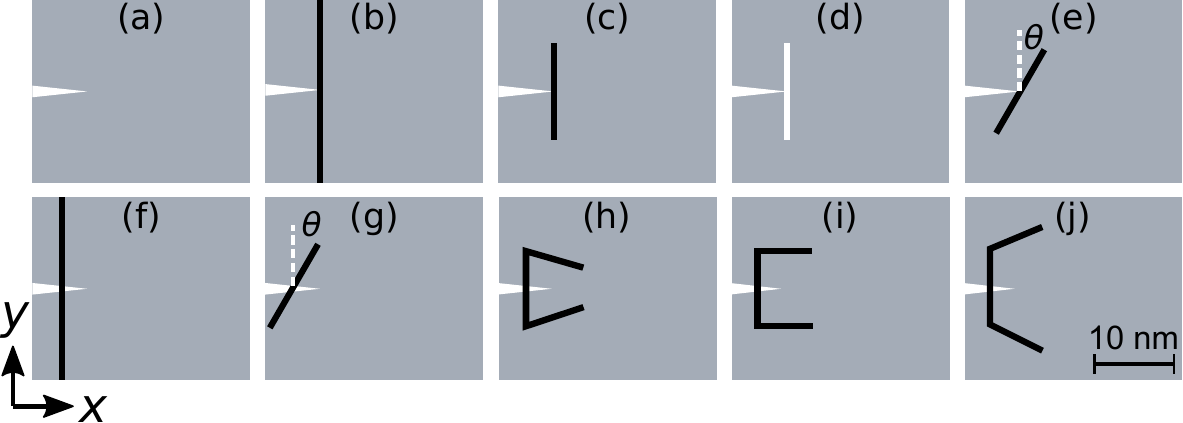}}
	\caption[Simulation cell set-up of crack/CNT interaction simulation]{Simulation set-ups for crack--CNT interactions (schematic). The crack surfaces are parallel to the xz-plane and the crack line runs the $z$ direction. (a) Al with an edge crack, (b-e) models for the effect of CNT on the crack deflection. In (b), (c) and (d) a long CNT, a short CNT, and a cylindrical cavity are located ahead of the crack tip, respectively. In (e) a short CNT is located ahead of the crack tip and makes an angle $\theta$ with the normal vector of the crack plane. (f-j) models for crack bridging by CNTs. In (f) a long CNT is located behind the crack tip. (g) an inclined short CNT is located behind the crack tip. (h-j) correspond to geometrically anchored CNT located behind the crack tip. (h) Al--CNT anchored by 60$\degree$ side branches, (i) Al--CNT anchored by 90$\degree$  side branches, (j) Al--CNT anchored by 120$\degree$ side branches.
}
	\label{fig:cr1-1}       
\end{figure*}

This section studies the interaction of cracks with CNT embedded into the Al matrix under tensile loading (mode I). We investigated the effects of CNT positions with respect to the crack tip, CNT angles to the normal vector of the crack plane, and geometries of entangled CNTs on the crack propagation and fracture toughness. Figure \ref{fig:cr1-1} schematically depicts models of Al-CNT systems with an edge crack. Figure \ref{fig:cr1-1}(b-e) are models for examination of the effect of CNT on the crack blunting (deflection), where a CNT is placed ahead of the crack tip. Figure \ref{fig:cr1-1}(h-j) are models to study the effect of CNT on the crack bridging, where a CNT is placed behind the crack tip. A pure Al matrix with an edge crack as shown in Figure \ref{fig:cr1-1}(a) is the reference model used for comparison.\par

\subsubsection{Sample preparation and simulation set-up}
\label{subsec:3.2.1}

We considere an Al single crystal with faces aligned with the [100], [010], and [001] directions of the fcc crystal lattice. Al atoms are initially located put on the sites of a perfect fcc lattice. The system size is 300.64 $\times$ 200.52 $\times$ 202.48 $\si{\angstrom}^3$ in $x$, $y$, $z$ directions, respectively. The created Al matrix is relaxed using the CG algorithm.  A sharp edge crack is introduced into the crystal by removing the interaction between the atoms on both sides of the crack plane as in \cite{Cui2014-msea}. The crack plane is the $xz$ plane and the crack line runs parallel to the $z$ direction. The crack length (=72.86 $\mathrm{\AA}$) is chosen to be sufficiently long to avoid any dislocation nucleation in the bulk before a crack extension or dislocation nucleation at the crack tip take place. Surface boundary conditions are imposed in the $x$ and $y$ directions, whereas periodic boundary conditions are applied in the $z$ direction (along the crack front). This model is an example of a brittle crack system with a [001](010) crack system which can not be contained within any $\left\{111\right\}$-glide plane and therefore, the natural dislocation in an fcc lattice cannot be emitted from the crack front \cite{Gumbsch1995-JMR}. A single-walled CNT(10,10) is embedded into the Al crystal with its longitudinal parallel to the [010] direction. The embedded CNTs have either equal length, indicated as "long CNT" or have a 60 $\si{\angstrom}$ shorter length than the Al matrix, which is denoted as "short CNT". The energy of the system is minimized as described in section \ref{subsec:3.1.1}.\par

The system is loaded in tension (mode I) in the $y$ direction as follows. We used a Cartesian coordinate system with its origin at the center point of the block. Velocities of $v_y$ and $-v_y$, constant in time, are imposed in $y$ direction on the atoms of the uppermost and lowermost (010) planes (boundary atoms), initially located at $y = \pm L_y/2$.  Al atoms between the lower and upper boundaries are assigned initial velocities, which vary linearly as functions of the $y$ coordinate, $v(y) = 2 v_y y/L_y$.  The strain rate  $\dot{\epsilon_{yy}}$ of the tensile test is thus given by $\dot{\epsilon}_{yy} = 2v_y/L_y$. The tensile deformation is performed at a constant strain rate of 0.001 $\mathrm{ps^{-1}}$. The temperature of the system is set to 0.1 K.

\subsection{Crack bridging by CNTs}
\label{subsec:3.2.2}

 \begin{figure*}
	\centering
	\resizebox{0.8\textwidth}{!}{
	\includegraphics{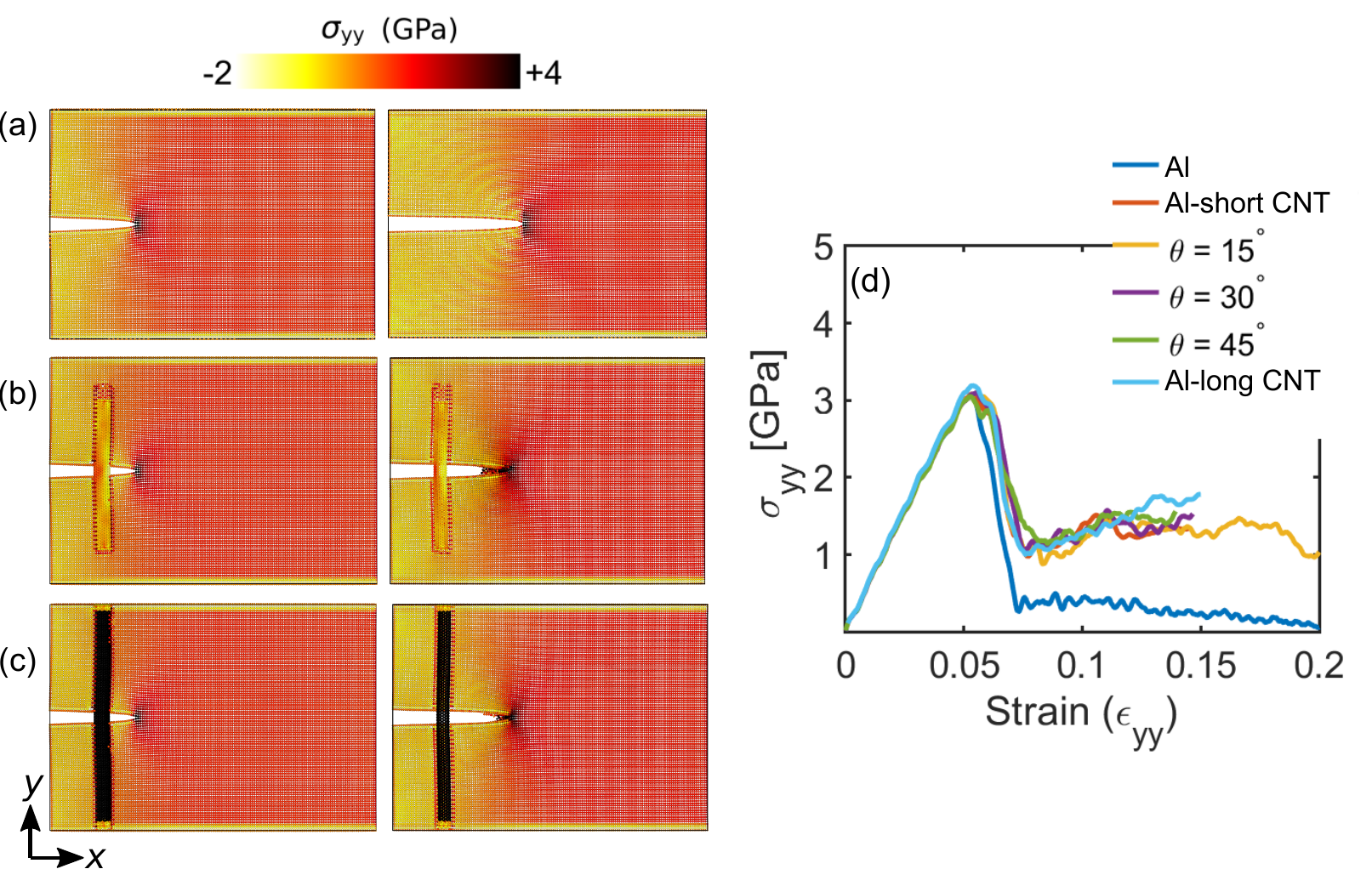}}
	\caption[Simulation of crack bridging in Al by CNT]{(a-c) Stress  patterns in pure Al, Al-short CNT, and Al-long CNT, respectively, (left) at the critical tensile strain and (right) after the onset of crack propagation. (d) Stress-strain curves for pure Al, Al-long CNT, and Al-short CNTs with different inclination angles ($\theta$).}
	\label{fig:cr2}       
\end{figure*}

Figure \ref{fig:cr2} (a-c) shows the edge crack conformation at (left), and after (right) the critical tensile strain, where the system reaches its maximum stress and the crack starts to propagate. Figure \ref{fig:cr2} (a) depicts a crack propagating in the pure Al matrix, whereas (b) shows a matrix reinforced with a short CNT and (c) with a long CNT. These snapshots represent sections along the $xy$ plane, which includes the CNT center axis, and atoms are colored based on the magnitude of their virial stress $\sigma_{yy}$ in the loading direction. As depicted in this figure, all systems show the same failure behavior. The crack starts to extend at a critical tensile displacement, corresponding to a sudden load drop in the stress-strain curves. The crack extension occurs at the critical strain $\epsilon^\mathrm{c}_\mathrm{Al} \approx 0.055$ in all simulations. The simulations of Al--short CNTs show that the CNT easily slides out of the Al matrix due to the low interfacial shear strength and, as a consequence, it carries little tensile load.  In the case of Al--long CNT, as the CNT is loaded geometrically parallel with the Al block, the CNT is under much higher stress compared to the short CNT, the long CNT does not affect the crack propagation process either, since there is again no shear stress transfer at the Al/CNT interface.  Hence, the presence of the CNT leaves the crack tip stress field and the crack face opening almost unchanged. The peak stress of the Al--long CNT composite is slightly higher than that of the pure Al and Al--short CNT, which is a direct consequence of the isostrain deformation. The inclination of the short CNT with respect to the crack plane normal has little influence on the stress-strain curves, as evidenced in Figure \ref{fig:cr2} (d).\par

\begin{figure*}
	\centering
	\resizebox{0.5\textwidth}{!}{
	\includegraphics{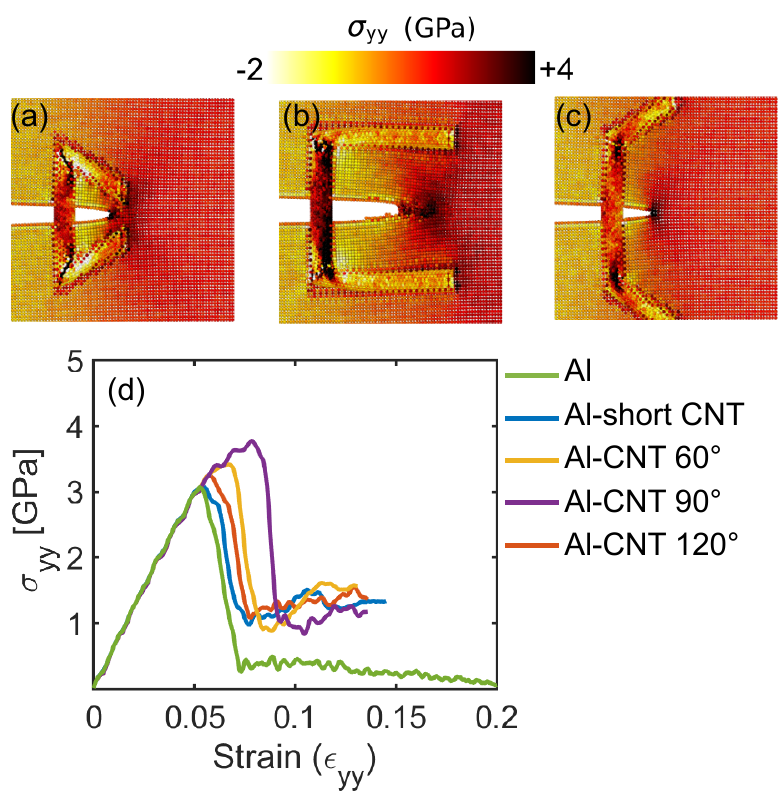}}
	\caption[Simulation of crack bridging in Al by geometrically anchored CNT]{Stress patterns around cracks bridged by CNTs that are anchored in the Al matrix by side branches under angles of (a) Al-CNT 60$\degree$, (b) Al-CNT 90$\degree$, and (c) Al-CNT 120$\degree$, and (d) corresponding stress-strain curves.}
	\label{fig:cr3}       
\end{figure*}
 
From experimental observations, it is known that CNTs are normally not isolated inside the matrix. Owing to their high surface area and energy, CNTs are usually found entangled with each other. This entanglement may anchor them in the Al matrix and delay the pullout process. Here we investigated the effect of geometrically anchored CNTs on the fracture process in terms of a few idealized initial configurations where CNTs are anchored in the matrix by inclined branches at angles of 60$\degree$, 90$\degree$, and 120$\degree$ as depicted in Figure \ref{fig:cr1-1} (h-j). The energy minimization steps during the sample preparation and the deformation procedure are the same as in the simulation of straight CNTs. Stress patterns for each model at the critical strain where the model systems reach their maximum stress, as well as the respective stress-strain curves, are shown in Figure \ref{fig:cr3}. \par

Upon loading, the anchored CNTs carry part of the tensile load, but more importantly, the presence of the side branches modifies the overall stress pattern such that additional stress concentrations emerge at the intersections between CNT and side branches, as well as at the side branches endings, see Figure \ref{fig:cr3} (a-c), where the black-colored regions represent the regions with a high magnitude of tensile stress. On the other hand, the crack tip stress concentration is reduced. This modification of the stress pattern leads to an increase in the overall stress and strain when crack propagation occurs. The effect is most pronounced for the CNT anchored by the side arms under an angle of 90$\degree$ where fracture stress and strain are increased by about 20\% and 35\%, respectively, compared to pure Al. \par

\subsection{Crack blunting by CNT}
\label{subsec:3.2.3}

This section investigates how CNTs can blunt a sharp, brittle edge crack in an aluminum matrix. We studied the effect of CNT length, and CNT inclination angle with respect to the normal vector of the crack plane. The initial configurations and deformation procedure are as explained in section \ref{subsec:3.2.1}. The only difference is that in this section, the CNTs are located ahead of the crack tip as depicted in Figure \ref{fig:cr1-1}(b-e). 

\begin{figure*}
	\centering
	\resizebox{0.5\textwidth}{!}{
	\includegraphics{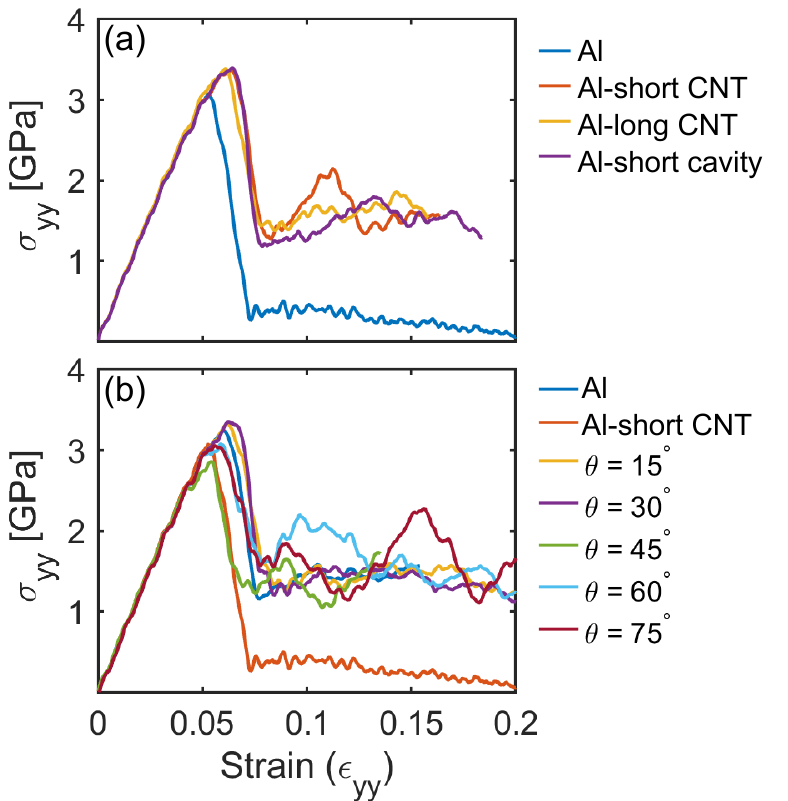}}
	\caption[Simulation of crack blunting in Al by CNT]{(a) Stress strain curves of a short and a long CNT, and a short cavity placed ahead of the crack tip in Al to model crack blunting. (b) Stress strain curves for short CNTs placed at the crack tip with various inclination angles ($\theta$).}
	\label{fig:cr4}       
\end{figure*}

The presence of a CNT located at the crack tip introduces multi-axial stresses that may facilitate dislocation nucleation and, as a consequence, induce plasticity and crack front blunting. We, therefore, analysed the effects of a CNT intersecting the crack tip. In order to understand whether these effects are due to the CNT or due to the cylindrical void in the Al lattice induced by the CNT, we also study the situation where the CNT is removed from its surrounding cavity. In this manner, we aim to clarify whether the CNT itself changes the fracture mode or the crack-front blunting is caused by the existence of a cylindrical void surrounding the CNT.\par
\begin{figure*}
	\centering
	\resizebox{0.9\textwidth}{!}{
	\includegraphics{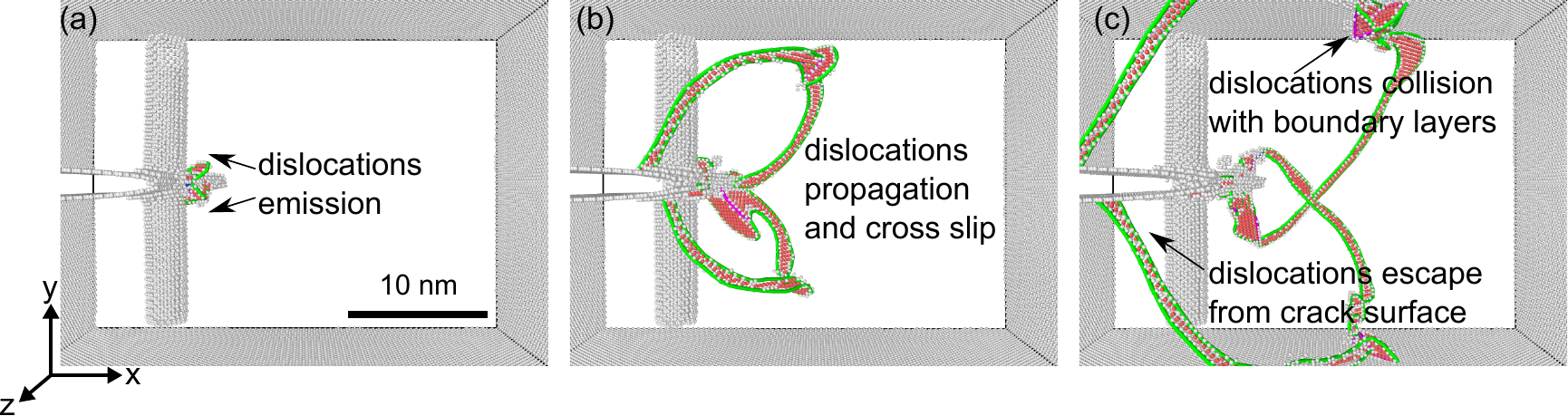}}
	\caption[Interaction of crack tip with CNT]{Deformation of Al--CNT composite under an applied tensile displacement in $y$ direction where a straight short CNT blocks the predefined edge crack. (a) $\epsilon_{yy} = 0.058$, (b) $\epsilon_{yy} = 0.062$, (c) $\epsilon_{yy} = 0.066$. Here only the atoms with perturbations of the fcc environment (icosahedral arrangement of nearest neighbors) are visualized. Green lines depict Shockley partial dislocations emitted from the crack tip. Pink lines show stair-rod dislocations. Atoms located at stacking faults are colored in red. Atoms with no crystalline structure are colored in light grey.}
	\label{fig:cr5}       
\end{figure*}

Figure \ref{fig:cr4} (a) shows that the Al matrix with a short CNT, a cavity, and a long CNT have in essence the same stress-strain curves. They all restrain the crack in a similar manner from expanding, leading to an increase of 10\% and 20\% in fracture stress and fracture strain, respectively. Figure \ref{fig:cr5} gives a closer look at the crack growth process in the case of a short CNT intersecting the crack tip. As the applied strain increases, small dislocation loops (which are split into Shockley partials) are emitted from the Al/CNT interface and expand on the crack surface as well as into the bulk. The CNT thus acts as a catalyst that facilitates dislocation nucleation and plastification of the Al matrix. This process is not contingent on the presence of the CNT - in fact, the cylindrical void surrounding the CNT {\em without} the embedded CNT acts in exactly the same manner.

\begin{figure*}
	\centering
	\resizebox{0.9\textwidth}{!}{\includegraphics{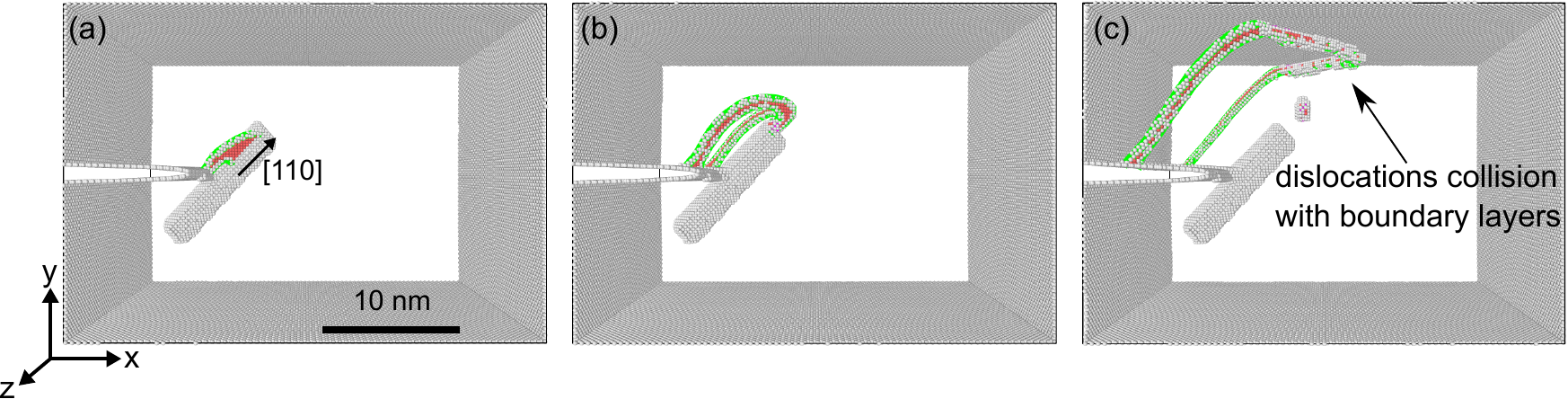}}
	\caption{Crack activity in Al where a $\theta$ = 45$\degree$ inclined CNT is placed at the crack-front, (a-c) $\epsilon_{yy}$ = 0.041, $\epsilon_{yy}$ = 0.043, and $\epsilon_{yy}$ = 0.046, respectively. The color scheme is same as in Figure \ref{fig:cr5}.}
	\label{fig:cr7}       
\end{figure*}

To understand the effect of CNT orientation, we placed a short CNT such that it intersects the crack tip and then rotates the CNT axis clockwise by steps of $\Delta\theta$ = 15$\degree$ around the direction of the crack front (Figure \ref{fig:cr1-1} (e)). The effect of CNT inclination angle $\theta $ on crack blunting is shown in Figure \ref{fig:cr4} (b). Stress concentrations at the Al/CNT interface and the intersection of CNTs' endings and Al facilitate dislocation nucleation, resulting in a combination of crack growth and dislocation emission from the Al/CNT interface. As a result of energy dissipation due to the ensuing dislocation motions, the stress-strain curves show toughening of the composite. This effect somewhat depends on the angle between the fracture surface normal and the CNT longitudinal axis. For orientations close to $\theta = 45\degree$,  the presence of the CNT can even have a slight weakening effect. 
To understand this effect, we looked closer at the deformation process of Al with a short CNT with the inclination angle of  $\theta $ = 45$\degree$, intersecting the crack tip as shown in Figure \ref{fig:cr7} from the side-view. In this case, the CNT is aligned with a Burgers vector of the fcc lattice (and accordingly with the intersection of two slip planes) and can, therefore, act as a very efficient site for dislocation nucleation from the Al/CNT interface, which happens before the critical fracture stress is reached. In Figure \ref{fig:cr7}, we can see that at a strain of 0.041, a pair of dislocation half loops of Burgers vector $<110>$ nucleate on both sides of the CNT at the locations where the CNT intersects the crack tip. These emergent loops split into Shockley partials and expand on the two conjugate (1-11) and (1-1-1) slip planes such that their endpoints travel on the crack surface and the Al/CNT interface.  Once they have reached the CNT endpoint, the two loops merge and detach from the CNT and propagate through the matrix until they collide with the fixed upper boundary layer or annihilate at the surface. 

\section{Conclusions}
We studied the deformation mechanisms of aluminum nanocomposites reinforced with carbon nanotubes of different sizes and geometries under tensile loading. We found that the physical nature of the interactions between Al and CNT significantly influences the composite's mechanical behavior. Van der Waals interactions between Al and $sp^2$ bonded carbon in CNT lead to weak bonding and a 'flat' interface energy landscape, which facilitates CNT pull-out from the surrounding Al matrix. Thus, uniaxial tensile simulations of Al-CNT, where CNTs are aligned parallel to the loading direction, showed that pristine CNTs do not provide significant reinforcement as such CNTs can easily slide inside the Al matrix. The same is true for CNT-crack interactions where crack bridging was found to be inefficient. 

A positive effect of CNTs is that their interfaces with the Al matrix may act as sites for dislocation nucleation at crack tips and thus provide enhanced toughness through crack blunting. However, this effect is not contingent on the presence of CNTs as simulations with the surrounding cylindrical void produce the same outcome. 

In view of the negative verdict on CNT reinforcement as far as load sharing and crack bridging are concerned, one may ask how the underlying root cause, namely weak interfacial bonding, can be mitigated. To this end, three different strategies are available: (i) The present study has demonstrated that CNTs that are geometrically entangled inside the Al matrix can act as efficient crack bridges and impede crack propagation. Such composites might be fabricated, e.g., by infiltration of self-entangled CNT networks \cite{schutt2017hierarchical}. (ii) To improve interfacial bonding, the chemical inertness of the CNT surface may be modified by chemical functionalization that offers sites for chemical bonding. In the context of Al-CNT composites, this might happen through carbide formation \cite{ci2006investigation}, which, however, needs to be carefully controlled to preserve the structural properties of the CNTs. (iii) An alternative which has been studied in the context of both metal-CNT and polymer-CNT composites consists in non-chemical functionalization of the CNT surface \cite{li2004carbon, Li2010-compos.sci.technol} which in the context of Al composites might be done by deposition of Ni nanoparticles or coatings \cite{Nasiri2019-EPJ} or surface decoration by other nanoparticles such as TiC \cite{saba2017formation}.
\par

\section*{Acknowledgments}
SN and MZ acknowledge financial support from DFG under grant no. Za171/11-1. The authors gratefully acknowledge the compute resources and support provided by the Erlangen Regional Computing Center (RRZE).


\section*{Conflict of Interest}
The authors declare that they have no known competing financial interests or personal relationships that could have appeared to influence the work reported in this paper.

\newpage		
\bibliographystyle{elsarticle-num} 
\bibliography{AlCNT_arXiv}

\end{document}